\documentclass[reprint,twocolumn,superscriptaddress,showpacs,nofootinbib,notitlepage]{revtex4-1}

\usepackage{graphicx}
\usepackage{latexsym,amsmath,amssymb,lmodern,float,url}
\usepackage{natbib}
\usepackage{color}
\usepackage{microtype}
\usepackage{slashed}
\usepackage{multirow}

\newcommand{\be}{\begin{equation}}
\newcommand{\ee}{\end{equation}}
\newcommand{\bea}{\begin{eqnarray}}
\newcommand{\eea}{\end{eqnarray}}
\def\beqs#1\eeqs{\be\begin{split} #1 \end{split}\ee}
\newcommand{\eq}[1]{Eq.~(\ref{eq:#1})}
\newcommand{\figu}[1]{Fig.~\ref{fig:#1}}

\def\tr {\mathop{\hbox{Tr}}}

\newcommand{\ba}{\begin{array}}
\newcommand{\ea}{\end{array}}

\newcommand{\re}{\mathop{\rm{Re}}}
\newcommand{\im}{\mathop{\rm{Im}}}
\newcommand{\as}{\langle\sigma\rangle}

\def\av#1{ \left\langle #1 \right\rangle }

\usepackage[colorlinks=true,backref=false, linktocpage=true,
citecolor=blue,urlcolor=blue,linkcolor=blue,pdfpagemode=UseOutlines]{hyperref}

\hypersetup{%
  bookmarksnumbered=true,
  pdftitle = {},
  pdfsubject = {},
  pdfauthor = {},
  pdfkeywords = {}
}

\def\Re {\mathop{\hbox{Re}}}
\def\Im {\mathop{\hbox{Im}}}

\renewcommand{\d}{\mathrm{d}}

\begin{document}
\title{Finite Density $QED_{1+1}$ Near Lefschetz Thimbles}

\author{Andrei Alexandru}
\email{aalexan@gwu.edu}
\affiliation{Department of Physics, The George Washington University, Washington, D.C. 20052, USA}
\affiliation{Department of Physics, University of Maryland, College Park, MD 20742, USA}
\affiliation{Albert Einstein Center for Fundamental Physics, Institute for Theoretical Physics, University of Bern, Sidlerstrasse 5, CH-3012 Bern, Switzerland}
\author{G\"ok\c ce Ba\c sar}
\email{gbasar@uic.edu}
\affiliation{Department of Physics, University of Illinois, Chicago, IL 60607, USA}
\author{Paulo F. Bedaque}
\email{bedaque@umd.edu}
\affiliation{Department of Physics, University of Maryland, College Park, MD 20742, USA}
\author{Henry Lamm}
\email{hlamm@umd.edu}
\affiliation{Department of Physics, University of Maryland, College Park, MD 20742, USA}
\author{Scott Lawrence}
\email{srl@umd.edu}
\affiliation{Department of Physics, University of Maryland, College Park, MD 20742, USA}
\date{\today}

\begin{abstract}
One strategy for reducing the sign problem in finite-density field theories is to deform the path integral contour from real to complex fields. If the deformed manifold is the appropriate combination of Lefschetz thimbles -- or somewhat close to them -- the sign problem is alleviated. Gauge theories lack a well-defined thimble decomposition, and therefore it is unclear how to carry out a generalized thimble method. In this paper we discuss some of the conceptual issues involved by applying this method to $QED_{1+1}$ at finite density, showing that the generalized thimble method yields correct results with less computational effort than standard methods.
\end{abstract}

\maketitle\section{Introduction}
Observables in non-perturbative field theory are predominantly calculated through stochastic methods and importance sampling. These techniques rely on the integrand of the path integral in  Euclidean space, $e^{-S}$, being real and non-negative, so that it can be interpreted as a probability. In several important cases, among them most theories at finite density, $S$ is complex, resulting in a non-real Boltzmann factor $e^{-S}$. Furthermore, $e^{-S}$ typically is a rapidly oscillating function of the field variables and computing the path integral becomes a challenge for any numerical method due to cancellations between different regions in field space. This issue is known as the {\em sign problem} and is the main roadblock to the lattice field theory study of dense quark matter (and several important problems in condensed matter physics).

In past years, a program to bypass the sign problem based on complexifying the field variables was set in motion. The path integral is evaluated, not as an integral over all real fields, but as an integral over a chosen integration contour in complex field space. Cauchy's theorem guarantees that, for a wide class of such contours, the value of the path integral is not altered by this procedure.
The original suggestion~\cite{Cristoforetti:2012su,Cristoforetti:2013wha,Cristoforetti:2013qaa,Scorzato:2015qts}  was to use a certain combination, $\mathcal{M}_{\mathcal{T}}$, of {\em Lefschetz thimbles}, the multidimensional analogue of the stationary phase paths of the complex function of one variable.  

A number of algorithms applied the idea of integrating over $\mathcal{M}_{\mathcal{T}}$ to toy models~\cite{Cristoforetti:2013wha,Fujii:2013sra,Alexandru:2016ejd,Alexandru:2017lqr} and fermionic field theories~\cite{Alexandru:2016ejd}.  Difficulties with this approach soon became apparent. First, it is difficult to determine the correct 
 combination of thimbles which is equivalent to the original path integral, a feat accomplished analytically only in some simple models~\cite{Tanizaki:2014tua,Kanazawa:2014qma,Fujii:2015bua,Tanizaki:2015rda}. Another obstacle is correctly sampling different thimbles with the appropriate weight~\cite{Alexandru:2017oyw,Fukuma:2017fjq}. For these and other reasons, new manifolds have been suggested. In~\cite{Alexandru:2015sua}, a so-called {\em generalized thimble method} (GTM) was proposed, in which a manifold is obtained by evolving the real space by the {\em holomorphic flow}.  If every point of $\mathbb{R}^N$, the original domain of integration, is evolved for a ``time" $T$, a manifold $\mathcal{M}_T$ is obtained that i) yields equivalent results to the original real space and ii) approaches $\mathcal{M}_{\mathcal{T}}$ in the limit $T\rightarrow\infty$, consequently improving the sign problem. However, for large flow times $T$, the parameterization of $\mathcal M_T$ by the real plane become ill-behaved. In particular, small regions on the real plane are mapped to large regions of $\mathcal M_T$, and most of the real plane is mapped to singular points of $\mathcal M_T$ which do not contribute to the integral. A Markov chain on the real plane is then unable to explore $\mathcal M_T$ effectively; this phenomenon is referred to as `trapping'. The GTM requires choosing $T$ large enough so the sign problem is sufficiently ameliorated but small enough that trapping is not an issue~\cite{Alexandru:2015sua,Alexandru:2016ejd,Nishimura:2017vav}.

Recently, more general manifolds have been used in order to cut down the computational cost of the complexification approach. For instance, a machine-learning technique was used to create a simple parametrization of a manifold $\mathcal{L}_{T}$ interpolating points obtained by the computationally expensive holomorphic flow~\cite{Alexandru:2017czx}. The holomorphic flow equations are completely avoided in another method proposed in~\cite{Mori:2017pne,Alexandru:2018fqp,Mori:2017nwj}, where $\mathcal{M}$ obtained by maximizing the average sign within families of manifolds.
 
 The application of these methods to gauge theories has developed more slowly, and only $0+1$ dimensional cases and one-plaquette models have been studied~\cite{Schmidt:2017gvu}. Gauge invariance brings a host of conceptual issues in the application of these methods. In this paper we applied the generalized thimble method to $QED_{1+1}$. Unlike in previously-studied models, the thimble decomposition in this model is not well-defined. This feature of gauge theories calls into question the use of the GTM, which is designed to approximate the thimble decomposition. In this paper, we show that the GTM formally yields correct results in this model. Additionally, we show how the definition of Lefschetz thimbles can be extended to be well-defined in an abelian gauge theory. Finally, we numerically demonstrate that the GTM both yields correct results, and improves the sign problem.
 
The paper is organized as follows. In Sec.~\ref{sec:gtm}, we briefly review the generalized thimble method, in Sec.~\ref{sec:model} we present the $QED_{1+1}$ model we study and discuss the particular features of it relevant to us. Sec.~\ref{sec:flow} presents the properties of the holomorphic flow in abelian gauge theories that insure correct answers. The results of the numerical calculations are presented in Sec.~\ref{sec:results} and further discussion is found in Sec.~\ref{sec:discussion}.
 
\section{Generalized thimble method}\label{sec:gtm}

We wish to calculate the thermal expectation value of an observable $\mathcal O$. In the path-integral formalism this is given by
\be\label{eq:expectation}
\left<\mathcal O\right> = \frac{\int \mathcal D\phi\;\mathcal O[\phi] e^{-S[\phi]}}{\int \mathcal D\phi\;e^{-S[\phi]}}\,,
\ee
where $S[\phi]$ is the Euclidean action. If $S$ is real, then the Boltzmann factor $e^{-S}$ is real and positive, and can be interpreted as a probability distribution. This expectation value can be efficiently approximated by importance sampling according to $e^{-S}$, and averaging the value of $\mathcal O$ over the collected samples:
\[
\left<\mathcal O\right> \approx \frac{1}{N} \sum_{i=1}^N \mathcal O[\phi_i]\,.
\]
In the limit $N\rightarrow \infty$, this approximation converges to the correct answer. It follows from the central limit theorem that the standard deviation of this estimator scales as $\delta_\mathcal{O} \sim N^{-1/2}$.

In this paper we are concerned with a fermionic field theory at finite chemical potential where the Euclidean action $S = S_R + i S_I$ is complex. In this case, the complex Boltzmann factor cannot be interpreted as a probability. The expectation value $\left<\mathcal O\right>$ must now be evaluated by reweighting:
\begin{align}
\left<\mathcal O\right>
&= \frac{\int\mathcal D\phi\;\mathcal O e^{-S}}{\int\mathcal D\phi\; e^{-S}}
\nonumber\\&= \frac{\left.\int\mathcal D\phi\;\mathcal O e^{-S_R} e^{-i S_I} \middle/ \int\mathcal D\phi\;e^{-S_R}\right.}{\left.\int\mathcal D\phi\; e^{-S_R} e^{-i S_I}\middle/\int\mathcal D\phi\;e^{-S_R} \right.}
\nonumber\\&= \frac{\left<\mathcal O e^{-i S_I}\right>_{S_R}}{\left<e^{-i S_I}\right>_{S_R}}\,,
\end{align}
where $\left<\mathcal O\right>_{S_R}$ is introduced to denote the expectation value taken with respect to the real part of $S$.
This estimator will still converge to the correct answer; however, the standard deviation of the estimator now scales with the average sign
$
\as \equiv \left<e^{-i S_I}\right>_{S_R}\,,
$: $\delta_\mathcal{O} \sim \left<\sigma\right>^{-1}N^{-1/2}$.  For fermionic theories in a finite volume $V$ and at finite density $n$, the average sign depends exponentially on $V$. Therefore, estimating a typical $\langle \mathcal O\rangle$ with a fixed precision requires a time exponential in $V$. This is the sign problem. In the remainder of this section, we describe a strategy that has been effective for ameliorating the sign problem in other models.

Our strategy is to deform $\mathbb R^N$ to $\mathcal M \subset \mathbb C^N$, where $\mathcal M$ is an N-real-dimensional manifold embedded in complexified field space on which the path integral remains unchanged but the sign fluctuations are milder. 
Cauchy's integral theorem guarantees that the integral over $\mathbb R^N$ will be the same as the integral over $\mathcal M$, as long as the integrand is holomorphic and the asymptotic behavior of $\mathcal M$ coincides with $\mathbb R^N$. As shown in Appendix~\ref{sec:cryptoholomorphy}, the integrands in the numerator and denominator of Eq.~(\ref{eq:expectation}) are both holomorphic for our theory; therefore we have
\be\label{eq:expectation-M}
\left<\mathcal O\right> ={\int_{{\mathbb R}^N}\mathcal D\phi\;\mathcal O e^{-S} \over \int_{{\mathbb R}^N}\mathcal D\phi\; e^{-S}} = \frac{\int_{\mathcal M} \mathcal D\phi\;\mathcal O[\phi] e^{-S[\phi]}}{\int_{\mathcal M} \mathcal D\phi\;e^{-S[\phi]}}\,.
\ee
How this deformation ameliorates the sign problem can be seen by examining what happens to the average sign:
\be\label{eq:averagesign}
\as\equiv \left<e^{-i S_I}\right>_{S_R} = \frac{\int \mathcal D \phi\; e^{-S}}{\int \mathcal D \phi\;e^{-S_R}}\,.
\ee
The numerator of $\as$ has a holomorphic integrand, and therefore is independent of $\mathcal{M}$. However, because the integrand of the denominator is \emph{not} holomorphic, $\as$ will take a different value on different manifolds. Integrating over an appropriately chosen $\mathcal M$ instead of $\mathbb R^N$ decreases the denominator, thereby alleviating the sign problem.

We now introduce a class of  manifolds, $\mathcal M_{T}$, obtained by continuously deforming $\mathbb R^N$ via the holomorphic flow equation:
\be\label{eq:flow}
\frac{\mathrm d \phi_i}{\mathrm dt} = \overline{\frac{\partial S}{\partial \phi_i}}\text.
\ee
Let $\tilde\phi_t(\phi)$ denote a solution to this equation, with initial conditions $\tilde\phi_0(\phi) = \phi \in \mathbb R^N$. The manifold $\mathcal M_T$ is defined by the collection of points $\tilde\phi_T(\phi)$ for all $\phi \in \mathbb R^N$. In other words, $\mathcal M_T$ consists of all points in $\mathbb R^N$ after flowing via Eq.~(\ref{eq:flow}) for fixed time $T$.  The flow time is a free parameter that we can tune in our computations to optimize total computational expense.

We parametrize every point $\tilde\phi_T(\phi)$ on $\mathcal M_T$ by the real coordinates in $\phi$.
Using the $\phi$ coordinates  the expectation value in Eq.~\ref{eq:expectation-M} can be written as
\begin{align}\label{eq:Oflow}
\left<\mathcal O\right>
&=
\frac{\int_{\mathcal M_T} \mathcal D\tilde\phi\;\mathcal O[\tilde\phi] e^{-S[\tilde\phi]}}{\int_{\mathcal M_T} \mathcal D\tilde\phi\;e^{-S[\tilde\phi]}}
\nonumber\\&=
\frac{\int_{\mathbb R^N} \mathcal D\phi\;\mathcal O[\tilde\phi_T(\phi)] e^{-S[\tilde\phi_T(\phi)]} \det J}{\int_{\mathbb R^N} \mathcal D\phi\;e^{-S[\tilde\phi_T(\phi)]}\det J}
\end{align}
where we introduce the Jacobian matrix given by
\be
J_{ij} =\frac{\partial \left(\tilde\phi_T\right)_i}{\partial\phi_j}\text.
\ee

In order to perform a Monte-Carlo computation of Eq.~(\ref{eq:Oflow}) we need a real, non-negative probability distribution. 
If we introduce $S_\text{eff}[\phi]\equiv S[\tilde\phi_T(\phi)]-\ln\det J$
a natural choice for this distribution is 
\be\label{eq:seff}
\left|e^{-S[\tilde\phi_T(\phi)]} \det J\right|= e^{-\re S_\text{eff}} \,.
\ee
The remaining phase, 
$\exp(-i\im S_\text{eff})$, is reweighted as before:
\begin{equation}
\av{O} = \frac{ \int_{\mathbb R^N} \mathcal D\phi\;\mathcal O[\tilde\phi_T(\phi)] e^{-S_\text{eff}}}
{\int_{\mathbb R^N} \mathcal D\phi\;e^{-S_\text{eff}}} = \frac{\av{\mathcal O e^{-i\im S_\text{eff}}}_{\Re S_\text{eff}}}{\av{e^{-i\im S_\text{eff}}}_{\Re S_\text{eff}}}
\end{equation}
where subscript means the importance sampling is performed with the probability distribution $e^{-\re S_\text{eff}}$ . 

We now explain briefly why deforming the original domain, $\mathbb R^N$, to $\mathcal M_T$ is likely to improve the sign problem. In the limit of $T\rightarrow\infty$, most fields acquire a large $S_R$ and decouple from the path integral due to their exponentially small Boltzmann weights, $e^{-S_R[\tilde \phi_T]}$. In terms of the parameterization by the real fields (where the Monte-Carlo sampling is performed), the support of the path integral comes from the points that flow into the critical points where the flow stops, and the small neighborhoods around them. These neighborhoods are stretched by flow and generate $N$-real-dimensional surfaces around the critical points. In other words, in the large T limit, $\mathcal M_T$ is the union of a set of approximate thimbles attached to these critical points. At the same time, $S_I$ is constant with flow, so the variation of $S_I$ on a given approximate thimble (in the real field parameterization) is small. Consequently the sampled fields has a small variation of the phase, hence a milder sign problem\footnote{Provided that the residual phase from $\im\det J$ is small, and cancellations \emph{between} thimbles are not large.}.

In order to compute $\det J$, the system of equations
\begin{equation}
\frac{d J_{ij}}{dt} = \overline H_{ik} \overline J_{kj},
\end{equation}
with $J_{ij}(t=0) = \delta_{ij}$ needs to be solved where the Hessian matrix is
\begin{equation}
H_{ij}(t) = \frac{\partial^2 S(\phi(t))}{\partial\phi_i \partial\phi_j}.
\end{equation}
This procedure is expensive, so we instead use an estimator $W$ in place of $\det J$, defined by~\cite{Alexandru:2016lsn}
\begin{equation}
W = {\rm exp}\int_0^T dt\  \tr\overline H(t).
\end{equation} The difference between $W$ and $\det J$ is reweighted when computing observables
\be
\av{ \mathcal{O}}
=
\frac
{ \langle
 \mathcal{O} e^{-\Delta S}
 \rangle_{S_\text{eff}'}  }
{\langle  e^{-\Delta S}
\rangle_{S_\text{eff}'} },
\ee 
where $S_\text{eff}' \equiv S - \ln W$, $\Delta S \equiv S_\text{eff} - \re S_\text{eff}'$, and $\av{\cdot}_{S_\text{eff}'}$ is
the average with respect to $\exp(-\re S_\text{eff}')$.
In this way, the expensive $\det J$ needs to be computed only for the configurations used for measurements, rather than for every step of the Monte Carlo chain.

\section{QED in 1+1 dimensions}\label{sec:model}
As a demonstration of the generalized thimble method in a gauge theory, we will use a three-flavor version of $QED$ in 1+1 dimensions. In the continuum, the Euclidean action is
\begin{equation}
	S = \int \d^2 x\; \left[ F_{\mu\nu}F^{\mu\nu} + \bar\psi^a \left(\slashed\partial + \mu_Q\gamma_0 + m - g Q_a \slashed A\right) \psi^a \right]
\end{equation}
where the field-strength tensor is $F_{\mu\nu} = \partial_\mu A_\nu - \partial_\nu A_\mu$, $m$ is the bare fermion mass, $g$ is the gauge coupling constant, the latin index $a$ sums over $N_F$ flavors, and $Q_a$ denotes the charge assigned to flavor $a$. Because $QED_{1+1}$ provides a long-range force, 
states with a net charge have a divergent energy in the thermodynamic limit and are excluded of the spectrum. In order to have a finite fermion density while maintaining charge neutrality we then need more than one flavor. The two flavor case with fermions with opposite charge (and equal mass) is not interesting for our purposes as the sign problem disappears. We use a model with three flavors with charge assignments $Q_1=Q_+=+2$, and $Q_{2,3}=Q_-=-1$, so states with a finite fermion number but zero charge are possible (for instance, a three fermion state with one fermion of each flavor).
We will refer to each fermion as a ``quark" and groups of three quarks, one of each flavor, as a ``baryon".
In this model, we take the quark chemical potential $\mu_Q$ to be the same for all flavors.

Discretizing Euclidean spacetime and integrating out the fermions yields a lattice action in which the only degrees of freedom are the bosonic vector field $A_\mu$, with two components at each site of the lattice. The action is
\begin{equation}
	S = \frac 1 {g^2} \sum_r \left(1 - \cos P_r\right) - \sum_a \ln \det D^{(a)}
\end{equation}
where $D^{(a)}$ denotes the fermionic matrix for flavor $a$, and $P_r$ denotes the primitive plaquette with $r$ at the lower-right corner:
\be
P_r \equiv A_1(r) + A_0(r+\hat x) - A_1(r+\hat t) - A_0(r)\,.
\ee
Above $\hat{t}$ and $\hat{x}$ are the unit vectors in time and space direction.
We discretize the fermion action using the staggered formulation.
The Kogut-Susskind staggered fermion matrix for flavor $a$ is given by
\beqs
	D^{(a)}_{xy} = &m_a \delta_{xy} + \frac 1 2 \sum_{\nu\in\{0,1\}}
	\big[
		\eta_\nu e^{i Q_a A_\nu(x) + \mu \delta_{\nu 0}}\delta_{x+\hat\nu,y} \\&\phantom{xxxxxxxxxx}- \eta_\nu e^{-i Q_a A_\nu(x) - \mu \delta_{\nu 0}} \delta_{x,y+\hat\nu}
	\big]
	\text.
\eeqs
Note that one staggered flavor corresponds in the continuum limit to either a pair of two-spinor-component fields or one four-spinor-component
field. We take all flavors to have the same mass $m_a = m$.
For our specific assignments of flavors and charges, the lattice action may be written
\begin{equation}
	S = \frac 1 {g^2} \sum_r \left(1 - \cos P_r\right) - \ln \det D_+ - 2 \ln \det D_-
\end{equation}
where $D_{+}$ ($D_{-}$) gives the fermion matrix for the flavor with charge $Q_+$ ($Q_{-}$).  This notation will be used in the following sections.

\section{Flow and Thimbles in an Abelian Gauge Theory}\label{sec:flow}

The integration over the manifold $\mathcal{M}_T$ yields correct physical expectations values: this, we reiterate, is guaranteed by Cauchy's theorem. However, it is not sufficient for the GTM to get the correct answer; in order to be of practical use, it must also improve the sign problem over the real plane calculation. Previous, heuristic arguments for indicating that GTM improves the sign problem relied upon $\mathcal{M}_T$ approaching the Lefschetz thimbles at $T\rightarrow\infty$.  Where no such unique thimble decomposition exists, it is not obvious what the large-$T$ limit of $\mathcal M_T$ is, and whether this manifold will have a reduced sign problem. 

In $QED_{1+1}$ dimensions, there are two separate obstacles to defining a unique thimble decomposition: critical points are degenerate (the action does not change along gauge orbits), and lines of flow may connect one critical point to another (Stokes' phenomenon). In this section we discuss these difficulties, showing that the Lefschetz thimble formalism can be changed to accommodate gauge orbits in general, and neither obstacle substantially adversely affects the performance of the GTM.

We look first at the difficulties presented by gauge orbits in configuration space.  A Lefschetz thimble is defined from an isolated critical point $z_c$ as the union of all solutions to \eq{flow} which asymptote to the critical point at large negative flow times: $\lim_{T\rightarrow -\infty} z(T) = z_c$. For a holomorphic function of $N$ complex variables, each isolated critical point is a saddle point with $N$ stable and $N$ unstable directions; therefore, the Lefschetz thimble defined in this way is an $N$-real-dimensional surface. The Lefschetz thimble decomposition is a union of a certain subset of these thimbles. However, in a gauge theory, all critical points can be continuously connected to infinitely many other critical points by gauge transformations, i.e., critical points form gauge orbits. Therefore, in order to use the Lefschetz thimble decomposition to understand how the GTM behaves when applied to a gauge theory, we must understand how the Lefschetz thimble formalism can be repaired in the presence of gauge orbits.

For $QED_{1+1}$, where the gauge group is abelian, the degeneracies introduced by gauge symmetry can be resolved in a straightforward fashion by gauge-fixing. Before proceeding, we stress that we will ultimately perform calculations \emph{without} gauge-fixing; gauge-fixing is merely a convenient tool to see how the Lefschetz thimble formalism may be repaired for an abelian gauge theory.
In the complexified field space a general gauge transformation is given by 
\begin{equation}
	A_{\mu}(x) \rightarrow A_{\mu}(x) + \alpha(x+\hat\mu) - \alpha(x)
\end{equation}
where $\alpha(x)$ is any complex-valued function on the lattice points and $\hat \mu$ denotes the unit vector along the direction $\mu \in \{0,1\}$. With $V$ lattice sites, $A_{x,\mu}$ can be represented as a $2V$ dimensional vector, so the complexified field space is $\mathbb C^{2V}$. The gauge orbit of any configuration $A_{x,\mu}$ is 
obtained by adding to it all vectors of the form $\alpha(x+\mu)-\alpha(x)$, for every $\alpha(x)$. This is a $(V-1)$-dimensional space since a constant $\alpha(x)=\alpha$ does not generate a gauge transformation. 
So the gauge orbits form a set of parallel $(V-1)$-dimensional linear subspaces, and each gauge-fixed slice is a $(V+1)$-dimensional subspace. Every $A_\mu(x)$ can be decomposed 
\be\label{eq:decomposition}
A_\mu(x)=A_\mu^\perp(x)+A_\mu^\parallel(x)\text,
\ee
with $A_\mu^\parallel(x)$ parallel to the gauge orbit and $A_\mu^\perp(x)$ 
orthogonal to it\footnote{The situation for non-abelian groups is more involved, and will be treated in a later work.}.
We can choose $A_\mu^\perp(x)$ as the representative of $A_\mu(x)$ in every gauge orbit. In this way, we decompose the original real configuration space as $\mathbb R^{2V} = \mathcal M_0 \oplus \mathcal G$, where $\mathcal M_0 = \mathbb R^{V+1}$ is the gauge-fixed space of $A_\mu^\perp(x)$, and $\mathcal G$ is a single gauge orbit.
The gradient $\overline{\frac{\partial S}{\partial A}}$ is orthogonal to the gauge orbits so
a gauge-fixed slice, defined by constant $A_\mu^\parallel(x)$, is invariant under the holomorphic flow.

In each gauge-fixed slice, critical points are now isolated, and there is no gauge-related obstruction to the definition of a Lefschetz thimble decomposition. Furthermore, the behavior of flow in this gauge theory can be understood by considering the behavior of flow in each gauge-fixed slice. Let $\mathcal M_T$ be the result of flowing $\mathcal M_0$ for some time $T$. Because the flow commutes with gauge fixing, the result of flowing the entire gauge-free integration domain is $\mathcal M_T\oplus \mathcal G$. In our simulations, we do not gauge fix, so that our simulations perform a random walk in the gauge orbit space $\mathcal G$ -- this has no effect since we only evaluate gauge-invariant quantities.

\begin{figure*}[t]
	\includegraphics[width=\linewidth]{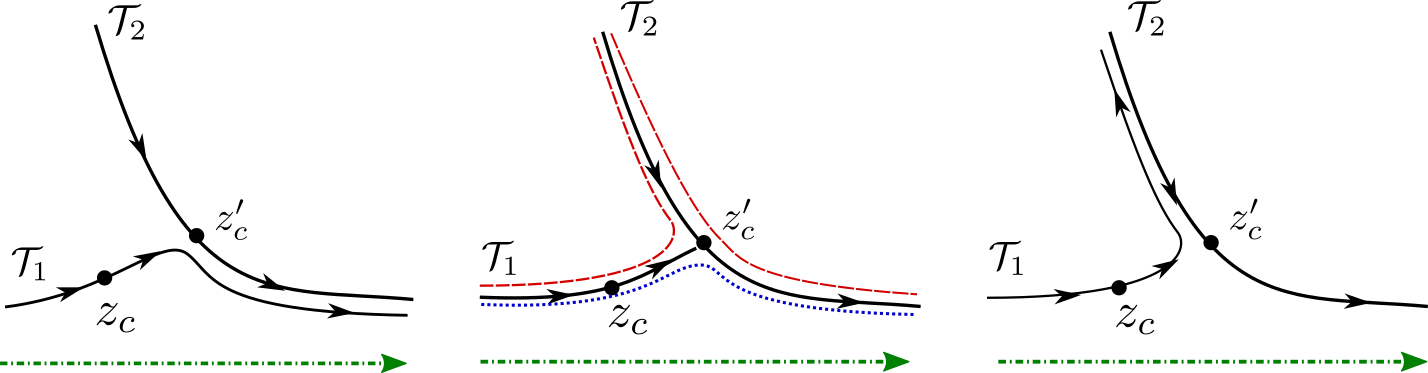}
	\caption{A schematic view of Stokes' phenomenon. The original integration contour is given by the dash-dotted green line. The solid dots denote the critical points, solid lines denote different thimbles, and the arrows denote the orientation of the thimbles. As the argument of $S$ changes (from left to right), the thimble decomposition jumps from $\mathcal T_1$ (left) to $\mathcal T_1$+$\mathcal T_2$ (right). When Stokes' phenomenon occurs (center) there is no unique thimble decomposition, which we indicate by two paths of integration given by the red dashed and blue dotted lines.
\label{fig:stokes}}
\end{figure*}

The second difficulty encountered in the thimble decomposition in $QED_{1+1}$ is Stokes' phenomenon, where a flow line connects two critical points. A schematic representation of Stokes' phenomenon is shown in the central panel of \figu{stokes}.  Stokes' phenomenon can be avoided by adding a small $i \epsilon$ to some parameter of the action. While the resulting thimble decomposition is well-defined, it differs depending on the sign of $\epsilon$.
The left and right panels of \figu{stokes} show the flow assuming such a modification of the action with opposite signs. In each case, the thimble decomposition is well-defined, but the combination of thimbles equivalent to the real domain is different depending on the sign of $\epsilon$.

\begin{figure}[b]
	\includegraphics[width=\linewidth]{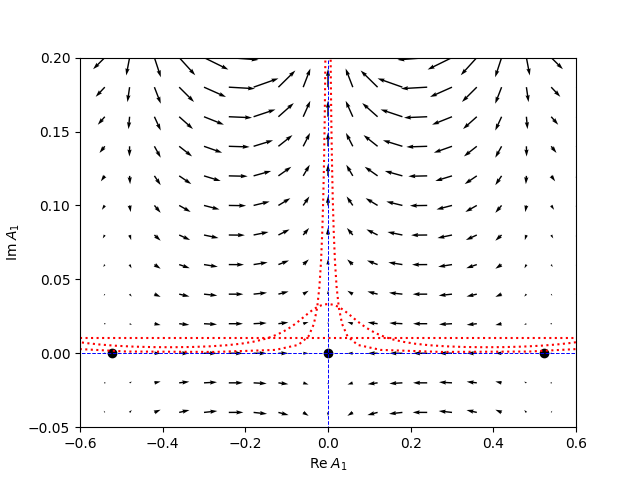}
	\caption{The $A_0 = 0$, $A_1 = \text{const.}$ complex plane for the case of $g=0.5$, and $\mu=0.1$.  Solid circles denote critical points. The vector field portrays the right-hand side of Eq.~(\ref{eq:flow}). The dashed blue lines show the stable thimbles: lines of flow emanating from shown critical points. The intersection of thimbles demonstrates Stokes' phenomenon. Red dotted lines show $\mathcal{M}_T$ for $T = 0.0,0.05,0.15$, generated by flowing from $\mathbb{R}+0.01i$.\label{fig:stokesqed}}
\end{figure}

The GTM does not rely on the thimble decomposition, but it is instructive to see how Stokes' phenomenon affects the behavior of the holomorphic flow. We investigate this particularly in the case of $QED_{1+1}$, for which Stokes' phenomenon occurs at all values of chemical potential $\mu$ and coupling $g$. 
For instance, consider fields with $A_0(x) = 0$ and $A_1(x)=A_1$ (constant). 
This particular slice of field space is depicted
 in \figu{stokesqed}.  In this slice, there are three critical points: one at $A_1=0$ and two others at non-zero values of $A_1$, both of which have flow lines going into the central critical point at $A_1 = 0$.  
%
Although Stokes' phenomenon prevents us from selecting $\mathcal{M}$ via a thimble decomposition, it is clear that integrating on a manifold $\mathcal{M}_T$ (red dashed lines) obtained by flowing from $\mathbb{R}^N$ yields results equal to those obtained on $\mathbb{R}^N$. The presence of Stokes' phenomenon does not cause any discontinuity in the flowed manifold.
On the other hand, Stokes' phenomenon does produce undesirable ``bumps'' in the deformed paths, depending on the orientation of the critical points and the starting manifold for the flow. In \figu{stokesqed}, several $\mathcal{M}_{T}$ are drawn, flowed from a shifted version of the real line $\mathcal M = \mathbb R + 0.01i$. As the amount of flow increases, a bump is created near the origin.

\begin{figure*}[t]
 \includegraphics[width=0.48\linewidth]{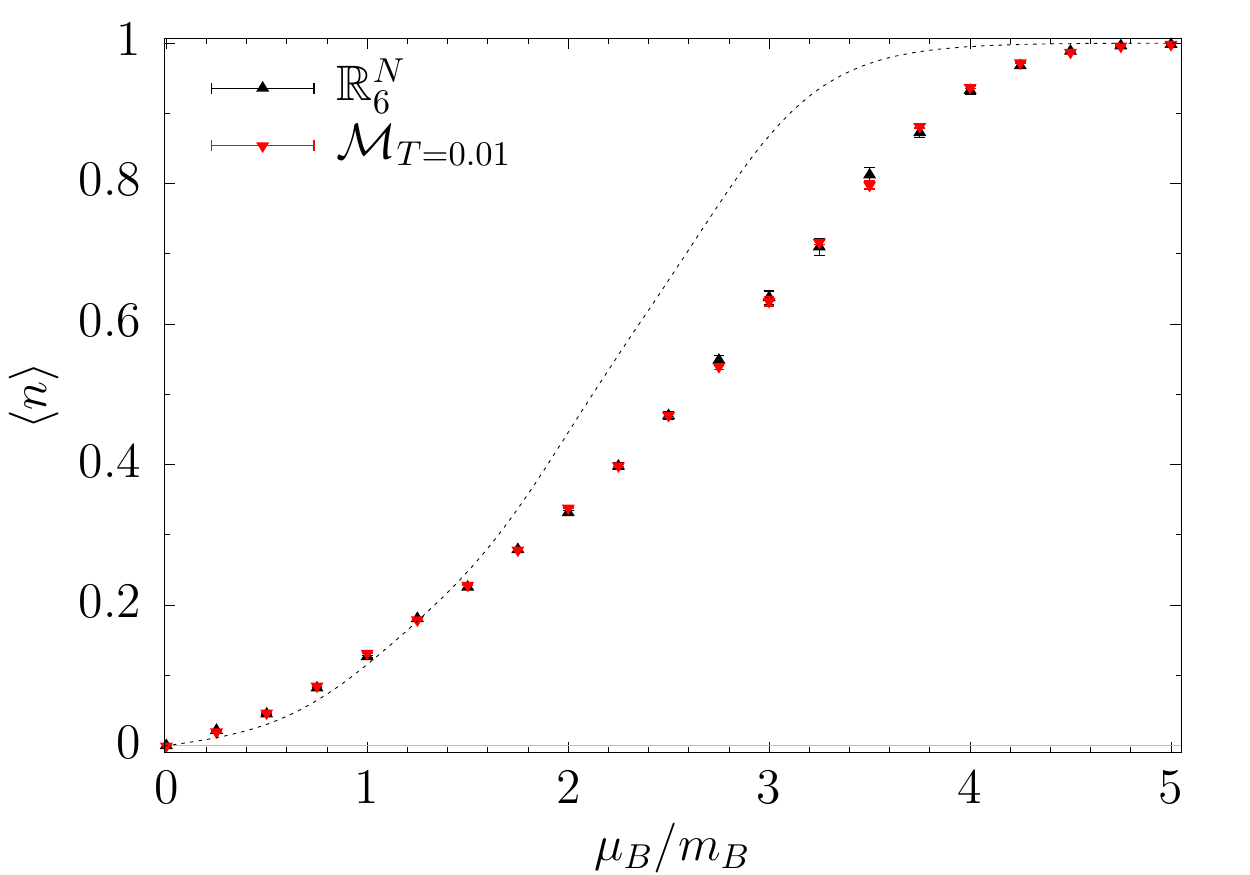}
 \includegraphics[width=0.48\linewidth]{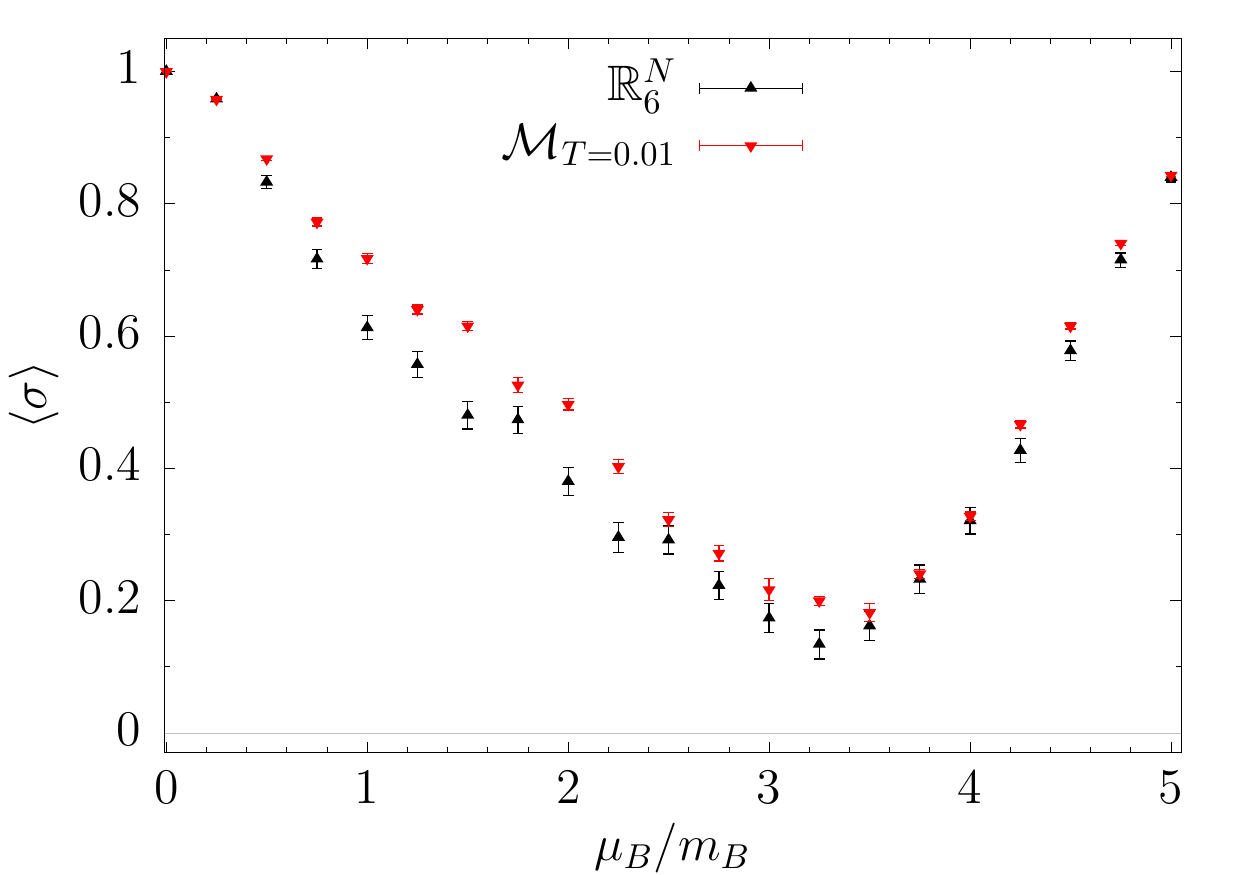}
\caption{\label{fig:610}Density $\langle n\rangle$ and average sign $\av{\sigma}$ as a function of $\mu_B/m_B$ for staggered fermions on a lattice of size $6\times10$.  The dashed curve represents the free baryon gas with the same mass.}
\end{figure*}

The effect of this bump is best understood by considering the $T\rightarrow\infty$ limit, in which it becomes the most pronounced. In this limit, the bump travels up one half of the imaginary axis, and directly back down again, producing a closed contour of integration. The sum of these two segments of the contour cancel and do not contribute to the integral. However, the average sign is decreased by the presence of such closed contours. The effect on the sign problem is bounded because the bump is produced along a path of steepest descent, so that the integral of $|e^{-S}|$ along the bump is finite.  This ensures that in the large $T$ limit, $\as$ asymptotes to a non-zero value.

\section{Results}\label{sec:results}

In this section we present some numerical results supporting the point that the GTM leads to the correct results -- and improves the sign problem -- even in gauge theories where the thimble decomposition is not well-defined.
We choose  the bare parameters $g=0.5$ and $m=0.05$ so that the renormalized baryon mass lies below the lattice cutoff scale.  We estimate the baryon mass $m_B$ by observing the onset of non-zero charged fermion density at low temperature (see the Silver Blaze discussion below).  The relation between the baryon chemical potential and the quark chemical potential allows us to find $a m_B\approx 0.6$. Since the ``baryon'' in our model is made up of three fermions we use $\mu_B=3\mu$ to quote the values of the 
chemical potential.

In this study, we have undertaken calculations on a fixed spatial lattice size at four different temperatures: $n_t\times n_x=6\times10$, $10\times10$, and $14\times10$ using both the GTM and a calculation done with real fields.  The parameters for the $\mathbb{R}^N$ and $\mathcal{M}$ simulations can be found in Table~\ref{tab:latdat}.

\begin{table}[b]
 \begin{center}
  \begin{tabular}
   {l| c c c c c}
   \hline\hline
   Lattice & $\mu_B/m_B$ & $T$ & $n_{\rm thermo}$ & $n_{\rm skip}$ & $n_{\rm config}$ \\
    \hline
    $6\times10$ &[0,5] &0 &48000 &480 &1000\\
    &[0,1] &0.01  &48000 &480 &5000-10000\\
    \hline
    $10\times10$ &[0,5] &0 &80000 &1200 &1000\\
    &[0,2] \& [3.5,5] &0.01  &80000 &1200 &500-12000\\
    &[2.5,3.25] &0.02  &80000 &2400 &1000-2000\\
    &0.60 &0.05  &40000 &2400 &1000\\
    \hline
    $14\times10$ &[0,3] &0 &28000 &2800 &500-12000\\
    &[0.75,4.75] & 0.01 &56000 &2800 &200-3000\\
    \hline
  \end{tabular}
     \caption{\label{tab:latdat}Metropolis sampling parameters.  $T$ is the flow time. Different values of $\mu_B/m_B$ generate flowed configurations at different rates, so for brevity we quote the number of configurations as ranges.  $n_{\rm thermo}$, $n_{\rm skip}$, and $n_{\rm config}$ are the thermalization, decorrelation lengths and number of measurements respectively in the Metropolis sampling.}
 \end{center}
\end{table}

To separate the effect of the phase fluctuations on $\mathcal M_T$ and the fluctuations induced by approximating $\ln\det J$ with $W$,
we write the reweighting factor $\Delta S = i\Im S_\text{eff} + \Delta J$.  
The phase factor $\exp(-i\Im S_\text{eff})$ is a pure phase and converges to the ``residual phase'' on the thimble as $T\to\infty$.
The real factor $\Delta J$, with $\exp(-\Delta J)=|\det J|/|W|$, is the reweighting necessary to correct for using $W$ instead of 
the Jacobian $\det J$ in the Monte-Carlo process.

In order to study the speed-up of the GTM  compared to real-plane calculations, one must consider: $t_\text{config}$, the wall-clock time required to generate a statistically independent configuration; $\langle\sigma\rangle=\av{\exp(-i\Im S_\text{eff})}$, the average sign; and $\Sigma$, the {\em statistical power} defined by  $w=e^{-\Delta J}/\langle e^{-\Delta J}\rangle_{S_{\rm eff}'}$.
The expression for $\Sigma$ is
\begin{equation}
 \Sigma\equiv\frac{\langle w\rangle_{S_{\rm eff}'}}{\langle w^2\rangle_{S_{\rm eff}'}}=\frac1{\av{w^2}_{S_\text{eff}'}}.
\end{equation}
$\Sigma$ is bounded between $1/n_\text{config}$ and $1$ and is an estimate of the fraction of configurations that effectively contribute to the statistics of the observable. A small value of $\Sigma$ indicates that the averages are dominated by a small number of configurations and, consequently, more configurations need to be used for a reliable estimate.

The reweighting is only necessary in the calculations on $\mathcal{M}_\mathcal{T}$, since $\Sigma$ is 1 when no flowing is done.  We find empirically that the statistical power tracks the average sign, decreasing with volume and $\mu$, although as explained below for the average sign, it recovers at saturation densities.  Unlike the average sign, $\Sigma$ tends to decrease with $T$ (although it asymptotes to a finite, non-zero value). As a consequence, the increase of flow time $T$ does not necessarily yield superior performance. 
  
\begin{figure*}[t]
 \includegraphics[width=0.48\linewidth]{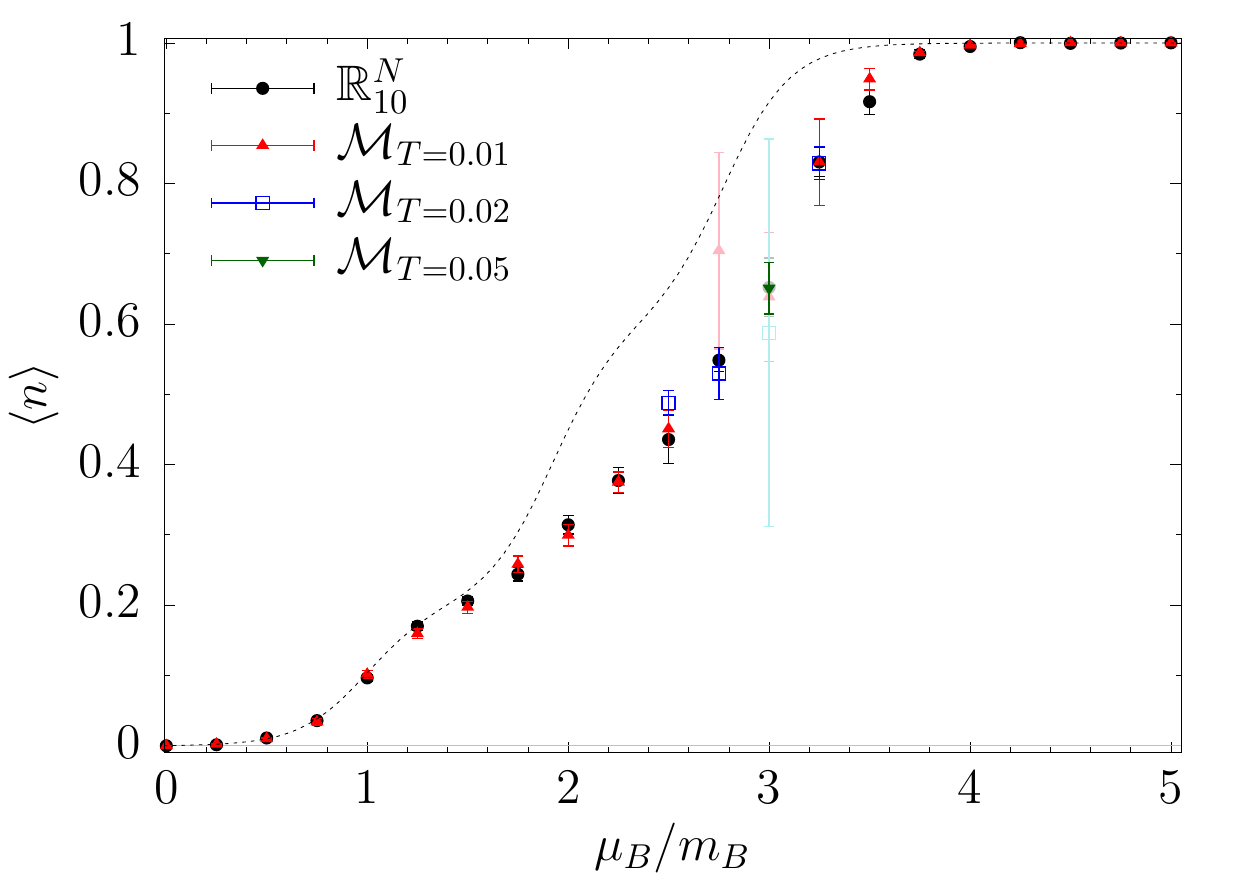}
 \includegraphics[width=0.48\linewidth]{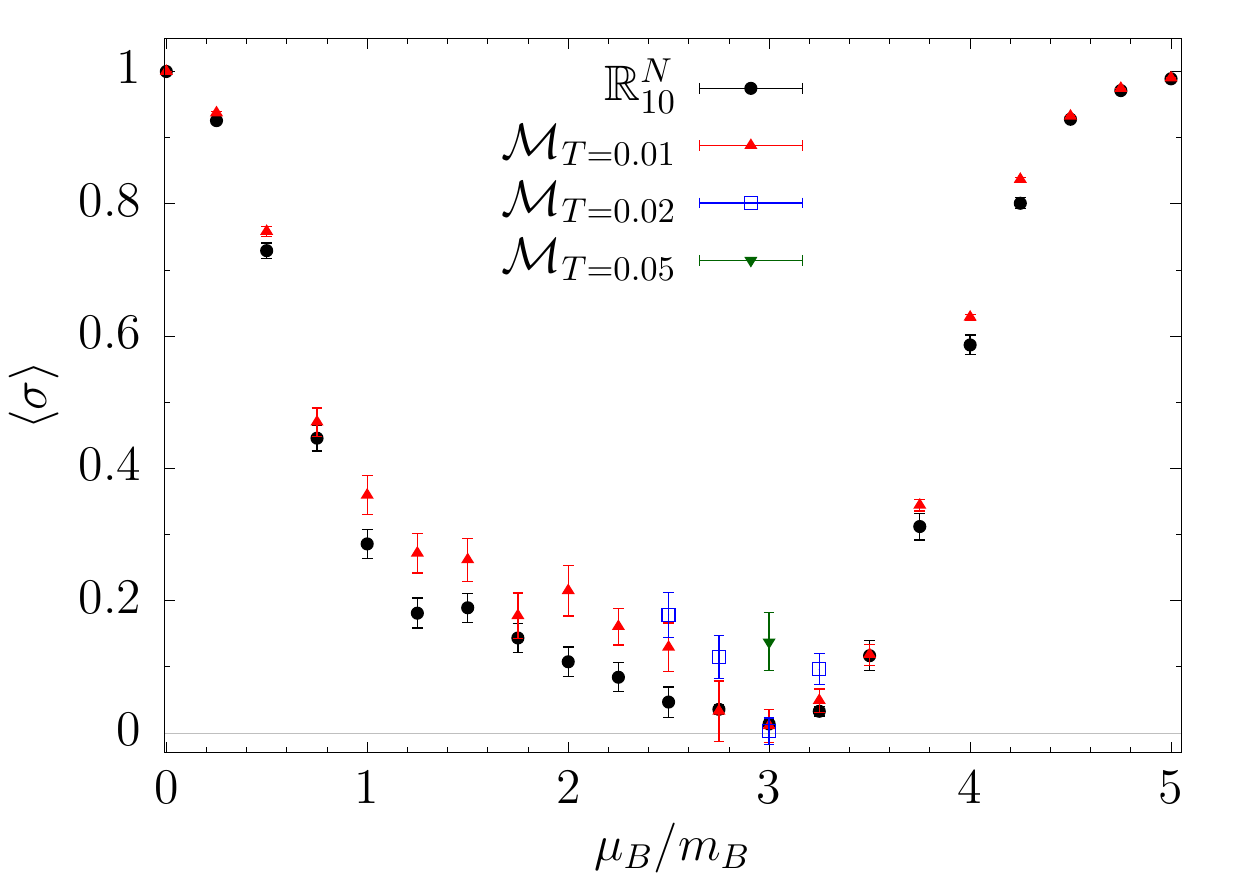}
\caption{\label{fig:1010}Density $\langle n\rangle$ and average sign $\av{\sigma}$ as a function of $\mu_B/m_B$ for staggered fermions on a lattice of size $10\times10$.  The dashed curve represents the free baryon gas with the same mass.}
\end{figure*}
  
We define a figure of merit $h_T$ for a fixed flow time $T$ as 
\begin{equation}
\label{eq:speed}
 h_T= \frac{\langle\sigma\rangle^2\Sigma}{t_\text{config}}\text.
\end{equation}
A ratio $h_{T_1} / h_{T_2}$ estimates the relative speed-up of flow time $T_1$ over flow time $T_2$, for a fixed desired precision.  We expect  $\langle\sigma\rangle$ to increase with $T$ because flowing improves the relative sign, but $t_{\rm config}$ also increases because longer flow requires more computational time.  Due to the use of $W$, an approximate Jacobian that is computationally faster, the statistical reweighting plays a nontrivial part in judging the speedup. 

As we present our results we will focus on the fermion density, as a representative physical observable, and the sign
average $\av{\sigma}=\av{\exp(-i\Im S_\text{eff})}$ as a characteristic of the sign problem on the flowed manifold. We
will compare our results for the density against the results from a free gas with the same mass as our ``baryon'' mass. This
model is expected to describe well the results at small density but as the number of particles increases the interactions will
modify the results significantly.

As a first test, we studied a $6\times10$ lattice where $\as$ is large enough on $\mathbb{R}^N$ that computations can be taken at all  $\mu_B/m_B$ up to lattice saturation. A small flow time $T=0.01$ was used to show that GTM can improve the sign and that the two methods  agree within the small error bars over the entire $\mu_B/m_B$ range as seen in Fig.~\ref{fig:610}.  This should be taken as an empirical confirmation that GTM respects all of the conditions upon complexification that are necessary to produce physically correct results despite the lack of a unique Lefschetz thimble decomposition.  On this small lattice, the average sign and statistical power are nearly the same for $\mathcal{M}_{T=0.01}$ and $\mathbb{R}^N$, but $t_{\rm config}$ is 6 times larger for the flowed manifold, therefore $h_{0.01}/h_0\approx1/6$, indicating that there is no computational benefit to using the GTM.

In contrast to other models (e.g. the 1+1 Thirring model of Ref.~\cite{Alexandru:2018fqp}), in $QED_{1+1}$ $\as$ appears to approach unity for large $\mu_B/m_B$, with $\as\rightarrow 1$ as $\mu_B/m_B \rightarrow \infty$. The absence of a sign problem at saturation densities can be understood directly from the fermionic part of the effective action. In the large-$\mu_B$ limit, the fermion determinant becomes $\det D_{\pm} \rightarrow \exp\left[\beta V (\mu + i Q_{\pm}  \av{A_0})\right]$. Thus, the product of all three fermion determinants, in this limit, is a positive real number $e^{3\beta V \mu}$, and there is no sign problem. This phenomenon has been previously noted in QCD~\cite{Schmidt:2017gvu}, and appears to be a general feature of gauge theories.

On the $10\times10$ lattice (seen in Fig.~\ref{fig:1010}), the average sign obtained with a flow time  $T=0.01$ is slightly improved compared to $\mathbb{R}^N$. For $\mu_B$ around $2.8 m_B$ the average sign on both manifolds become indistinguishable from zero.  For those values of $\mu_B/m_B$, we increase $T$.  For all but $\mu_B/m_B\approx3$, $T=0.02$ is sufficient to discriminate the sign from zero, that is $\av{\sigma}>2\epsilon_\sigma$ where $\epsilon_\sigma$ is the statistical error on $\sigma$. For $\mu_B/m_B=3$ we  increase the flow time $T$ further, taking advantage of the fact that $\Sigma$ will asymptote to a fixed value (so that flow time can be increased without a penalty to the statistical power).  The results for $h_T/h_0$ can be seen in Table~\ref{tab:ht}. We find that for this lattice, flowed manifolds can reduce computational time compared to the real plane for some values of $\mu_B/m_B$.

\begin{figure*}
 \includegraphics[width=0.48\linewidth]{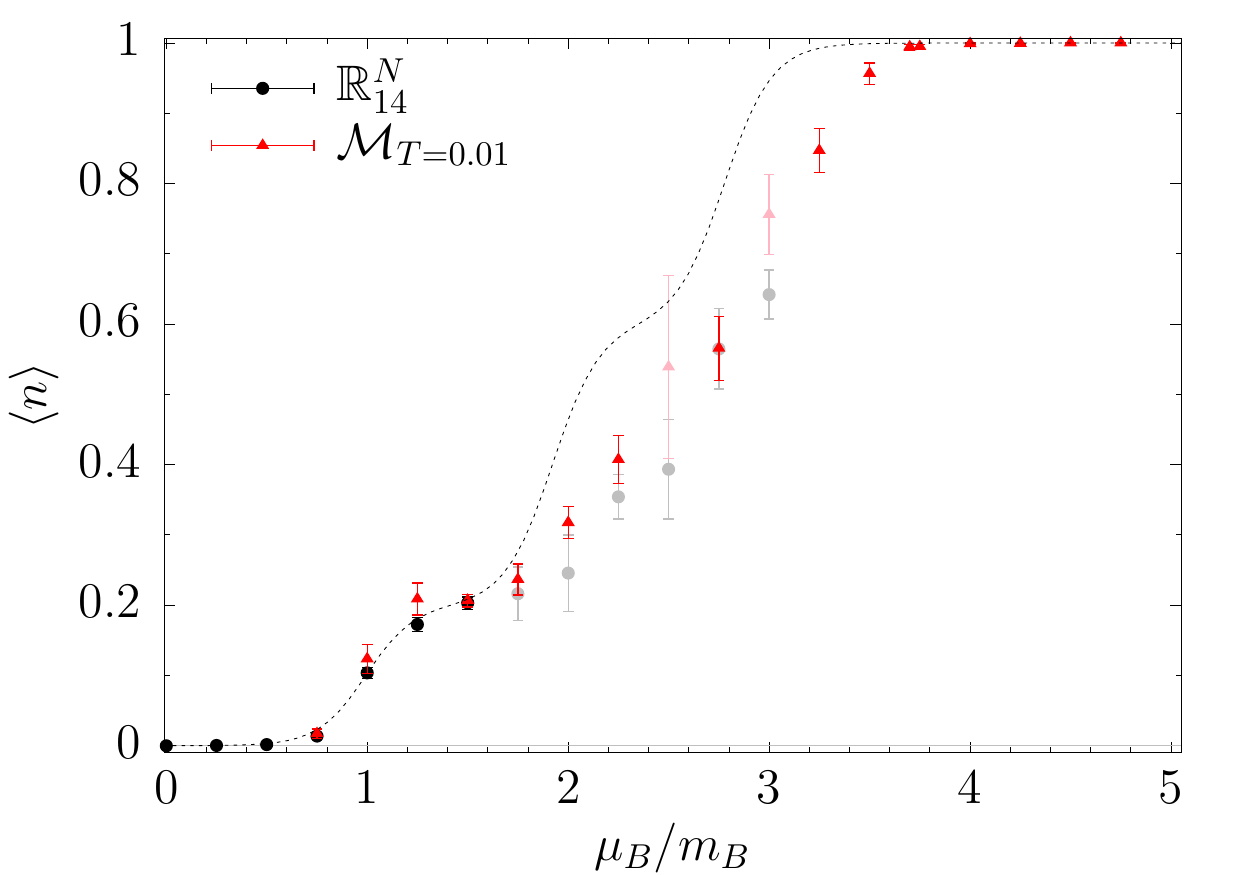}
 \includegraphics[width=0.48\linewidth]{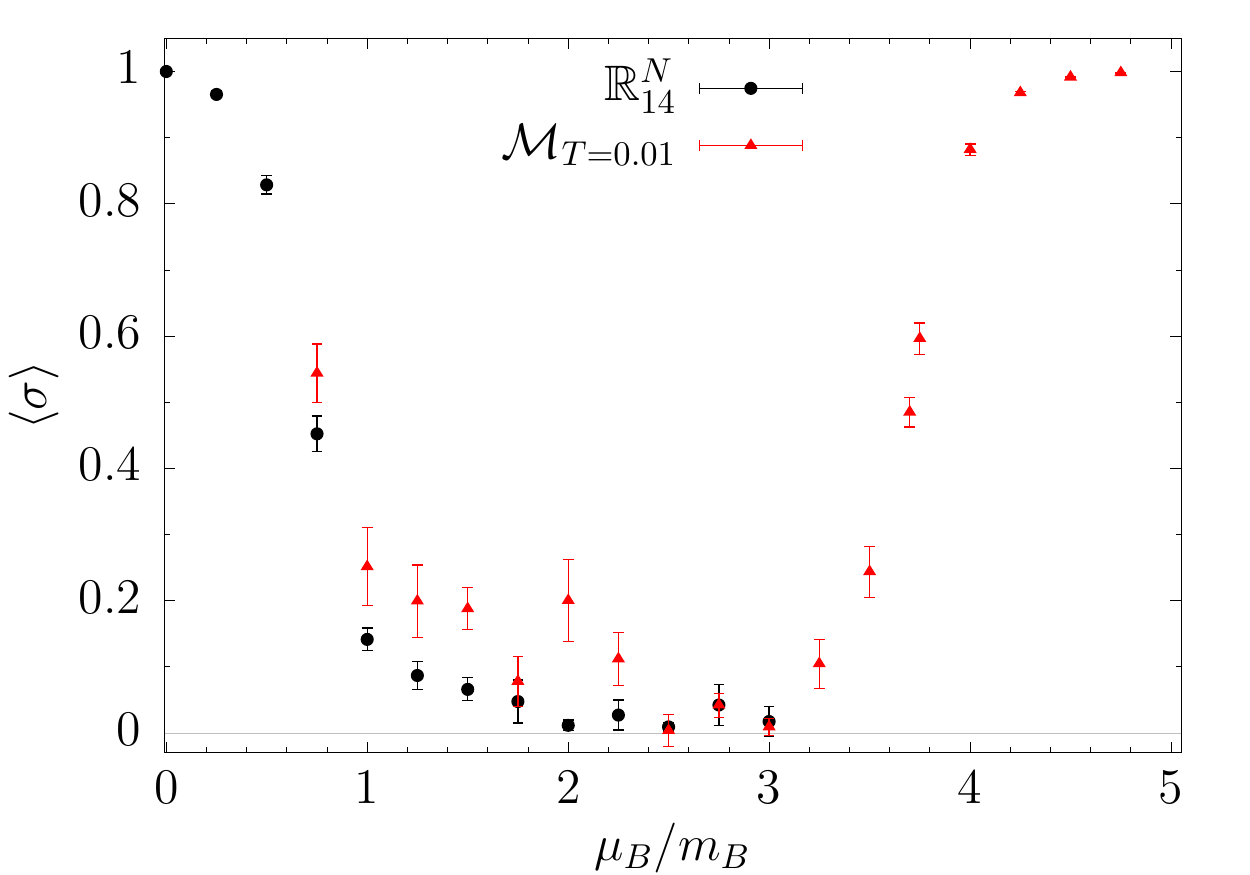}
\caption{\label{fig:1410}Density $\langle n\rangle$ and average sign $\av{\sigma}$ as a function of $\mu_B/m_B$ for staggered fermions on a lattice of size $14\times10$.  The dashed curve represents the free baryon gas with the same mass.}
\end{figure*}

\begin{table}[b]
 \begin{center}
  \begin{tabular}
   {l| c c c c}
   \hline\hline
   $T$ & $t_{\rm config,T}/t_{\rm config,0}$ & $\Sigma_T/\Sigma_0$ & $\langle\sigma\rangle_T/\langle\sigma\rangle_0$ & $h_T/h_0$  \\\hline
   0.01 & 6&0.8&3&1.2\\
   0.02 & 18&0.7&4&0.6\\
   0.05 & 28&0.5&13&3\\
    \hline
    \hline
  \end{tabular}
    \caption{\label{tab:ht}Maximum figure of merit $h_T/h_0$ for different flow times $T$ measured on the $10\times10$ lattice.}
 \end{center}
\end{table}

On our coldest lattice, 14$\times$10, one can see the onset of {\em Silver Blaze} phenomenon at small $\mu_B/m_B$ in Fig.~\ref{fig:1410}, as well as development of a plateau at the one baryon threshold.  For this lattice, $t_{\rm config}$ for $T=0.01$ is 7 times that of the real plane. We find for different $\mu_B/m_B$, $\Sigma_T$ ranges from 0.65 to 1 and $h_{0.01}/h_0$ ranges from  0.15 to 28.  Since the figure of merit for certain values of $\mu_B/m_B$ again exceeds unity, holomorphic flow is seen to reduce the computational time required to achieve a fixed-precision result at $\mu_B/m_B\approx 1.5$ .

\section{Discussion and prospects}\label{sec:discussion}

Working with $QED_{1+1}$, we have shown that the generalized thimble method may be applied to gauge theories without encountering any fundamental obstacles. None of the difficulties that plague the definition of Lefschetz thimbles in a gauge theory need to be confronted despite the fact that the original motivation for the GTM explicitly invoked the thimble decomposition.  Additionally, despite the holomorphic flow requiring greater computational time to produce a single configuration compared to the real plane, we have shown that the improvement in the average sign can reduce the total time needed to compute observables at a fixed precision to less then an equivalent real time calculation. 

Although the holomorphic flow gives a relative speed up over the real plane for $QED_{1+1}$, the computational cost is still large.  Extending this work to larger lattices, higher dimensions, and non-abelian theories will require the development of faster algorithms to deal with the Jacobian that do not suffer from small statistical power as the approximate Jacobian does at large flow times and larger lattices.

\appendix
\section{Holomorphic integrands}\label{sec:cryptoholomorphy}

The applicability of the generalized thimble method requires that the
integrands $e^{-S[\phi]}$ and $\mathcal O[\phi] e^{-S[\phi]}$ both be
entire (i.e., everywhere holomorphic) functions of the field configuration $\phi$. Critically, this does not
require that $\mathcal O[\phi]$ and $S[\phi]$ are themselves holomorphic. Indeed,
for the model considered here, neither is, because a fermion matrix $D$ has logarithmic singularities where $\det D = 0$. At the same points, $D^{-1}$ is not well-defined, so the
expectation value of a meson propagator (for instance) involves a singular
$\mathcal O[\phi]$:
\[
\left<\bar\psi_i \bar\psi_j \psi_j \psi_i\right> = \frac 1 Z \int \mathcal D\phi\; e^{-S[\phi]} \left[D^{-1}_{ij} D^{-1}_{ji} - D^{-1}_{ii} D^{-1}_{jj}\right]
\]
 In
this appendix we will establish that although $S[\phi]$ is not entire for $QED_{1+1}$, and $\mathcal O[\phi]$
is not entire for many observables, the integrands
$e^{-S[\phi]}$ and $\mathcal O[\phi] e^{-S[\phi]}$ are always holomorphic (with lattice regularized actions).

For simplicity, we will consider an action involving only a single species of
fermions, and therefore only one fermion determinant $\det D$. All arguments
given here generalize easily to the cases of two or more differently-charged
species. That $e^{-S[\phi]}$ is entire is easily established: by
definition, this integrand may be written
\[
e^{-S[\phi]} = e^{-S_B[\phi]} \det D[\phi]
\]
where the `bosonic' action $S_B$ is manifestly holomorphic. Each component
$D_{ij}$ of the fermion matrix is a holomorphic function of the gauge fields; since the determinant
is a polynomial in those components, $\det D$ is a holomorphic
function of $\phi$ as well.

To see that integrands involving fermionic observables are entire, we write an expectation value in terms of the original, fermionic path integral.
\[
\left<\bar\psi_a\psi_b\right> = \frac 1 Z \int \mathcal D\phi \;e^{-S_B[\phi]} \int \mathcal D\bar\psi \;\mathcal D \psi\; \bar\psi_a \psi_b e^{-\bar\psi_i D_{ij}[\phi] \psi_j}
\]
With $N$ sites, the fermionic exponential $e^{-\bar\psi D\psi}$ may be expanded
in $2^{2N}$ terms: each Grassman variable may be included in a term or not, and there are $2N$ Grassman variables. The commuting part of each term is a product of finitely many
components of $D$, and therefore is a holomorphic function of the fields
$\phi$. Multiplying by any combination of $\bar\psi\psi$ and integrating over
$\mathcal D\bar\psi\mathcal D\psi$ has the effect of selecting one of these
coefficients. Therefore, the integral over fermionic fields yields a
holomorphic function of $\phi$.

\begin{acknowledgments}
A.A. is supported in part by the National Science Foundation CAREER grant PHY-1151648 and by U.S. Department of Energy grant DE-FG02-95ER40907. A.A. gratefully acknowledges the hospitality of the Physics Departments at the Universities of Maryland and Kentucky, and the Albert Einstein Center at the University of Bern where part of this work was carried out. GB is supported by the U.S. Department
of  Energy  under  Contract  No.  DE  FG02-01ER41195.   
P.F.B., H.L., and S.L.  are supported by U.S. Department of Energy under Contract No. DE-FG02-93ER-40762.

\end{acknowledgments}
\bibliographystyle{apsrev4-1}
\bibliography{thimbology}
\end{document}